\documentclass[11pt]{article}
\usepackage{graphics}
\def\DZero {\mathrm{D0\!\!\!/ \ }}
\def\DZeroBf {\mathbf{D0\!\!\!/ \ }}
\def\EtMiss  {E_T\!\!\!\!\!\!/ \quad}
\title{Top Quark Cross-Section Measurements at the Tevatron}

\author{Wolfgang Wagner \\ 
  Universit\"at Karlsruhe \\ 
  Institut f\"ur Experimentelle Kernphysik \\ 
  Wolfgang-Gaede-Stra{\ss}e 1, 76128 Karlsruhe \\  
  \vspace*{4mm} \\
  for the CDF Collaboration \\ \vspace*{3mm}% etc
}                     % Do not remove

\begin{document}
\maketitle
\abstract{Run II of the Tevatron collider at Fermilab is well under way 
and data samples 
larger than those of Run I are at hand. In this contribution I summarize the
current status of cross-section measurements for top-quark pair ($t\bar{t}$)
production at the CDF and $\DZero$ experiments. This article is a
contribution to the proceedings of the EPS-HEP conference 2003 in
Aachen, Germany.
%
%\PACS{
%      {14.65.Ha}{Top quarks}   \and
%      {13.85.Lg}{Total cross sections}
%     } % end of PACS codes
} %end of abstract
%
%
%\title{\vspace*{-20mm}\\ {\normalfont CDF/PUB/TOP/PUBLIC/6740} \\ {\normalfont Version 1.0} \\ 
%\vspace*{5mm} \\
%Top Quark Cross-Section Measurements at the Tevatron}
%\subtitle{Do you have a subtitle?\\ If so, write it here}
%
%\offprints{Wolfgang Wagner, \\ wagner@ekp.physik.uni-karlsruhe.de} 
% Insert a name or remove this line
%
%
%\date{Received: date / Revised version: date}
% The correct dates will be entered by Springer
%
%
%
\section{Introduction}
\label{intro}
The Tevatron accelerator complex at Fermilab underwent a major upgrade
to prepare for Run II. 
Among other measures the beam energy was increased from 800 to 980~GeV, 
which leads to higher cross-sections for the production of heavy particles. 
The $t\bar{t}$ cross-section is expected to increase by about 30\%. 
Several other changes allow higher beam currents and thereby lead to a higher 
instantaneous luminosity. The aim for Run II is
to achieve $\mathcal{L}=5 - 20\cdot10^{31}\,\mathrm{cm^{-2}s^{-1}}$.
The highest luminosity reached to date is $5.2\,\mathrm{cm^{-2}s^{-1}}$.
Until early September 2003 the Tevatron experiments had recorded about 
240~$\mathrm{pb^{-1}}$.

In preparation for Run II both collider detectors were upgraded as well. 
New silicon trackers were installed. $\DZero$ added a solenoid in 
the bore of the detector to facilitate momentum measurements for
charged particles. The main tracking chambers were replaced.
CDF installed a new drift chamber, $\DZero$ a scintillating fiber
tracker. CDF replaced its plug calorimeter and extended its
muon coverage. Both experiments completely renewed their 
trigger and data acquisition systems.

The top physics program at the Tevatron comprises an exciting variety
of topics. While the $t\bar{t}$ cross-section measurements test perturbative
QCD, many other analysis investigate the nature of the top quark:
The top mass measurement is an important input to standard model fits
and constraining the mass of the Higgs boson. 
Measurements of the helicity of $W$ bosons from top quark decays test the
$V-A$ structure of the charged current. The search for anomalous top quark
decays such as $t\rightarrow u/c + \gamma / Z / g$ allow to look for
flavor-changing neutral currents and new physics. Searches for top
decays to a charged Higgs boson, $t\rightarrow b + H^+$ allow to
test SUSY models. While this article is concerned with the $t\bar{t}$ 
cross-section measurement, several of these topics are covered by
other contributions to these proceedings~\cite{maruyama,merkel,shabalina,warsinsky}.  

\section{Top Quark Pair Production}
\label{sec:1}
At the Tevatron top quarks are predominantly produced as quark-antiquark pairs
via the strong interaction. About 80 to 90\% of the $t\bar{t}$ cross-section
are due the quark-anti\-quark annihilation process $q\bar{q}\rightarrow t\bar{t}$,
the remaining share is due to gluon-gluon fusion.
Due to its large width ($\Gamma\simeq 1.4$~GeV) the top quark does not hadronize, 
but rather decays as a free quark. In the standard model top quarks decay 
almost entirely in the mode $t\rightarrow W^+ + b$ (BR $\simeq$ 100\%). 
This is an underlying assumption for all cross-section measurements discussed 
in this article.

Experimentally $t\bar{t}$ events are classified according to the subsequent
decays of the two $W$ bosons coming from the top quarks.
The $W$ decay modes are: $e\nu$, $\mu\nu$, $\tau\nu$ and $q_1\bar{q_2}$.
If both $W$ bosons decay into light leptons ($e\nu$ or $\mu\nu$) 
the $t\bar{t}$ event is called a {\it di-lepton} event
(4/81 of all $t\bar{t}$ events). 
If one $W$ decays to $e\nu$ or $\mu\nu$ and the other $W$ decays
into two quarks, the event belongs to the category {\it lepton-plus-jets} 
(24/81 of $t\bar{t}$ events).
The case where both $W$'s decay hadronically is called {\it all-hadronic}
channel (36/81). 
The remaining events fall into the category where at least one $W$ decays 
to $\tau\nu$ (17/81).

In Run I ($\sqrt{s}=1.80$~TeV)
CDF and $\DZero$ have measured the $t\bar{t}$ cross-section
in the di-lepton, the lepton-plus-jets and the all-hadronic channel.
The combined cross-section of all channels is $6.5^{+1.7}_{-1.4}$~pb for CDF 
and $(5.9\pm1.6)$~pb for $\DZero$.
In the following paragraphs I will discuss preliminary results
of cross-section measurements in the di-lepton and the lepton-plus-jets
channel for Run II.

\section{Di-lepton Channel}
\label{sec:2}
\begin{figure}[th!]
\begin{center}
% Use the relevant command for your figure-insertion program
% to insert the figure file.
% For example, with the option graphics use
\resizebox{0.7\textwidth}{!}{%
  \includegraphics{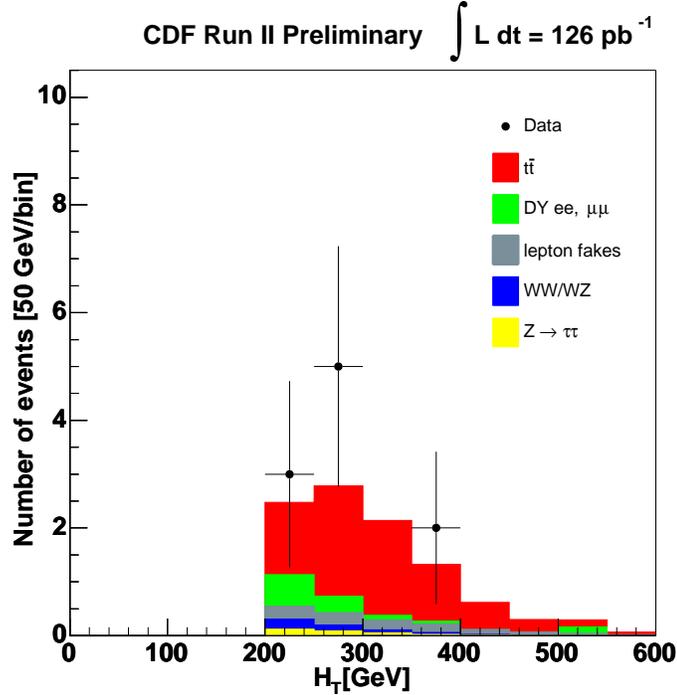}
}
\end{center}
% If not, use
%\vspace{5cm}       % Give the correct figure height in cm
\caption{$H_T$ distribution for CDF data and the standard model
expectation. $6.3\pm0.7$ $t\bar{t}$ events are expected.}
\label{fig:HTdilepton}       % Give a unique label
\end{figure}

The di-lepton $t\bar{t}$ decay mode features a fairly clean signature
and a low background level, 
however it suffers statistically from a small branching ratio.
The classical di-lepton analysis at CDF requires two oppositely
charged leptons ($e$ or $\mu$). The leptons have to pass standard
identification cuts and have a $p_t > 20$~GeV.
One of the leptons is required to be isolated.
The analysis uses leptons measured in the central and the
forward (plug) detectors.
If the di-lepton invariant mass falls in the $Z$ window
($76\,\mathrm{GeV}<m_{\ell\ell}<106\,\mathrm{GeV}$) additional cuts
are applied: (1) $\Delta \phi$ between the $\vec{\EtMiss}$ vector
and nearest jet has to be larger than $10^\circ$.
(2) The jet-significance 
\[ jetsig = \frac{|\vec{\EtMiss}|}{\sum \vec{E_T}(jet) \cdot \vec{\EtMiss}}\]
has to be larger than 8.
Neutrinos are detected by asking $|\EtMiss|>25\,\mathrm{GeV}$.
If $|\EtMiss|<50\,\mathrm{GeV}$, an additional requirement is
that the angle $\Delta \phi$ between the $\vec{\EtMiss}$ vector
and the nearest jet or lepton be larger than $20^\circ$.
Furthermore, the analysis calls for at least 2 jets 
($E_T>15\,\mathrm{GeV}$, $|\eta|<2.5$) and
$H_T>200\,\mathrm{GeV}$. $H_T$ is the scalar sum of all
transverse energies: 
$H_T=|\EtMiss|+|E_T(\ell_1)|+|E_T(\ell_2)|+\sum |E_T(jet)|$.
A veto to conversions and cosmic rays is applied.
CDF finds 10 candidate events (2 $e/e$, 5 $e/\mu$ and 3 $\mu/\mu$) 
passing all cuts in 125~$\mathrm{pb^{-1}}$ of data.
The expected number of background events is 
$2.9\pm0.9$. The deduced cross-section is
$\sigma(t\bar{t}) = 7.6 \pm 3.4\,(\mathrm{stat})\pm 1.5\,(\mathrm{syst})$~pb.
Fig.~\ref{fig:HTdilepton} shows the $H_T$ distribution for data in
comparison to the standard model expectation. 

CDF has performed a second, alternative di-lepton analysis
which is based on one tight lepton and one isolated track
(without additional lepton identification).
This analysis yields a cross-section of
$\sigma(t\bar{t}) = 7.4\pm3.4\,(\mathrm{stat.})\pm 1.7\,(\mathrm{syst.})$~pb,
which is in good a agreement with the standard analysis described
above.

$\DZero$ has observed 7 candidate events (4 $e/e$, 1 $e/\mu$ and 2 $\mu/\mu$)
in the di-lepton channel. The background expectation is $1.7\pm 0.6$. 
The resulting cross-section is 
$\sigma(t\bar{t}) = 29.9^{+21.0}_{-15.7}(\mathrm{stat})\,
 ^{+14.1}_{-6.1}(\mathrm{syst})\pm 3.0\,\mathrm{(lum)\;pb}$.

\section{Lepton-plus-Jets Channel}
\label{sec:3}
The background in the lepton-plus-jets channel is much more significant
than in the di-lepton mode. 
Dedicated methods to significantly reduce the backgrounds have to be applied.
There are three general strategies:
\begin{enumerate}
  \item Apply {\it topological cuts (or fits)} to kinematic or event shape
    variables, such as $H_T$ or aplanarity $\mathcal{A}$.
  \item Identification of {\it b-jets} based on {\it lifetime information},
    either by requiring a secondary vertex from a b-hadron decay
    or by selecting events that have tracks with large impact parameter
    significance.  
  \item Identification of {\it b-jets} by searching for 
    {\it non-isolated leptons from semileptonic b-decays} 
    (Soft Muon/Elec\-tron Tagger).
\end{enumerate}
In the following paragraph we will discuss the topological $t\bar{t}$
analysis at $\DZero$, the secondary vertex analysis of CDF and
the soft-muon-tagger analysis of $\DZero$.

\subsection{Topological $\mathbf{t\bar{t}}$ analysis at $\DZeroBf$}
The preselection of the topological analysis proceeds
by asking for an electron or a muon with $p_T>20$~GeV
and $\EtMiss > 20$~GeV. A veto is applied to soft muons, in
order to render a disjoint data sample with the b-tagged 
analysis (using the soft muon tagger).

\begin{figure}[th]
\begin{center}
\resizebox{0.8\textwidth}{!}{%
  \includegraphics{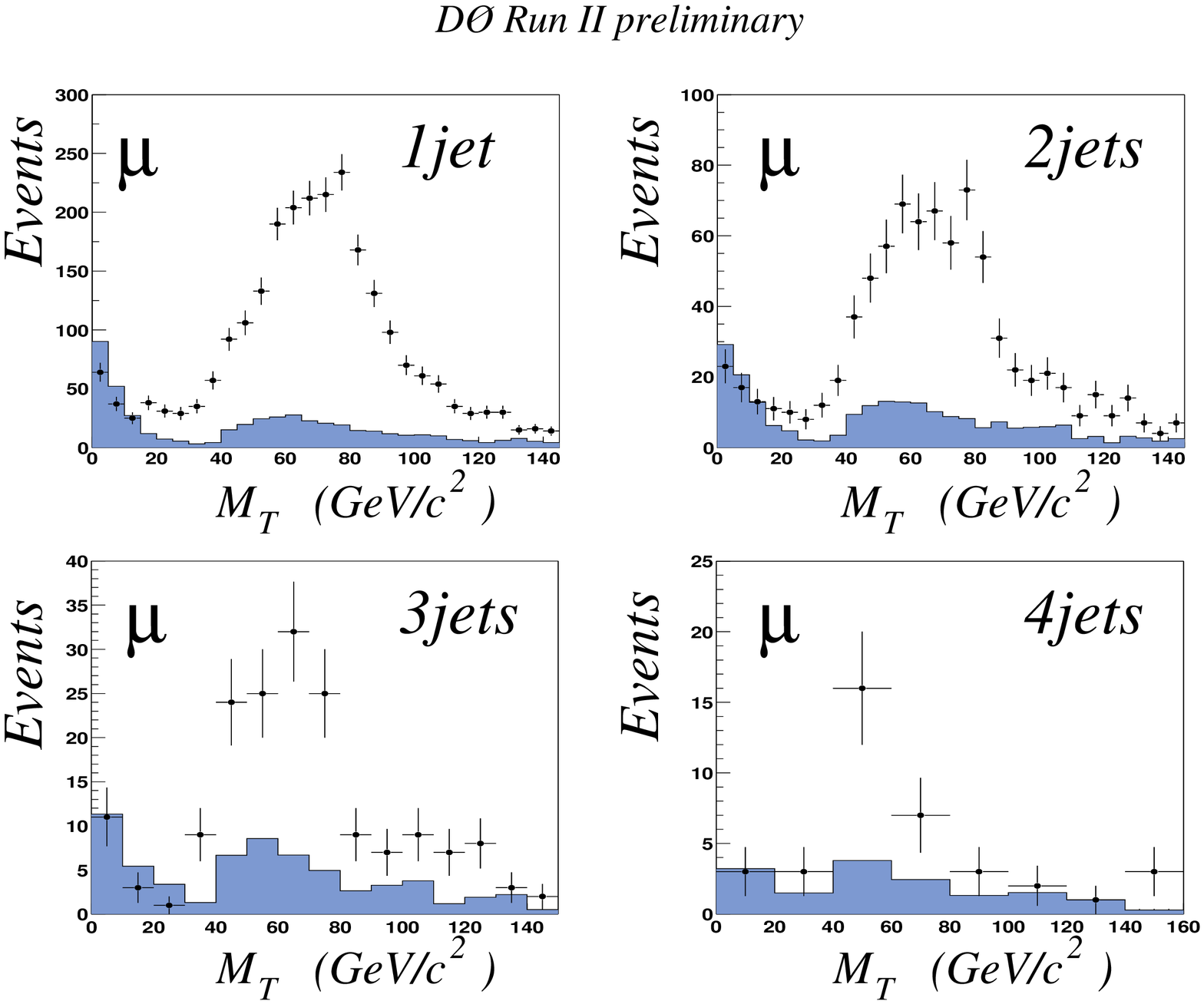}
}
\end{center}
\caption{Transverse mass spectrum of reconstructed $W$ bosons for different 
  jet bins in the muon channel. The multi-jet background is shown as 
  shaded histogram, the data are the dots with error bars.  
}
\label{fig:D0multijet}       % Give a unique label
\end{figure}

The preselected sample is used to estimate the backgrounds:
(1) the multi-jet background and (2) the $W$+jet background.
The multi-jet background is determined by applying an additional
isolation cut to the lepton. The cut efficiencies for multi-jet (QCD) events
and $W$+jet events are obtained from control samples.
Solving the two equations
\begin{eqnarray*}
  N_{obs} & = & N_{MJT} + N_{W+J} \\ \nonumber
  N_{iso} & = & \epsilon_{MJT} N_{MJT} + \epsilon_{W+J} N_{W+J} \nonumber
\end{eqnarray*}
gives the number of multi-jet background events $N_{MJT}$ in the
sample. Fig.~\ref{fig:D0multijet} depicts the transverse mass spectrum
of reconstructed $W$ bosons for different jet bins in the muon
channel. 
The multi-jet background is indicated as shaded histogram.
The $W$+jets background estimate is based on the empirical
Berends scaling law which predicts:
\[ \frac{\sigma \left( W + (n+1) \ \mathrm{jets} \right)}
        {\sigma \left( W + n \ \mathrm{jets} \right)} = \alpha \]
First the multi-jet background is subtracted from the data,
then $\alpha$ is obtained from a fit to the N-jet 
spectrum (using N=1,2,3). The fit predicts the number of
$W$+4 jets events. 
The $W$+4 jets bin is taken as the signal region in this 
top quark analysis.

The last step is to apply the topological cuts.
For electron events the analysis demands:
aplanarity $\mathcal{A}>0.065$ and $\sum E_T\,(\mathrm{jet}) > 180$~GeV.
For muon events: 
$\mathcal{A}>0.0065$ and $H_T > 220$~GeV.
$\DZero$ observes 4 electron+jets and 4 muon+jets events.
The number of expected background events is 
$1.8\pm0.4$ and $2.4\pm0.4$, respectively.
The resulting cross-sections are
\[\sigma(t\bar{t}) = 5.2^{+9.4}_{-6.7}(\mathrm{stat})\,^{+9.1}_{-3.1}
(\mathrm{sys})\pm0.52(\mathrm{lumi})\;\mathrm{pb}\] 
for the electron channel and
\[\sigma(t\bar{t}) = 3.8^{+6.9}_{-4.9}(\mathrm{stat})\,^{+3.9}_{-5.4}
 (\mathrm{sys})\pm0.38(\mathrm{lumi})\;\mathrm{pb}\]
for the muon channel.

\subsection{Secondary vertex based cross-section measurement at CDF}
\begin{figure}[thb]
\begin{center}
% Use the relevant command for your figure-insertion program
% to insert the figure file.
% For example, with the option graphics use
\resizebox{0.8\textwidth}{!}{%
  \includegraphics{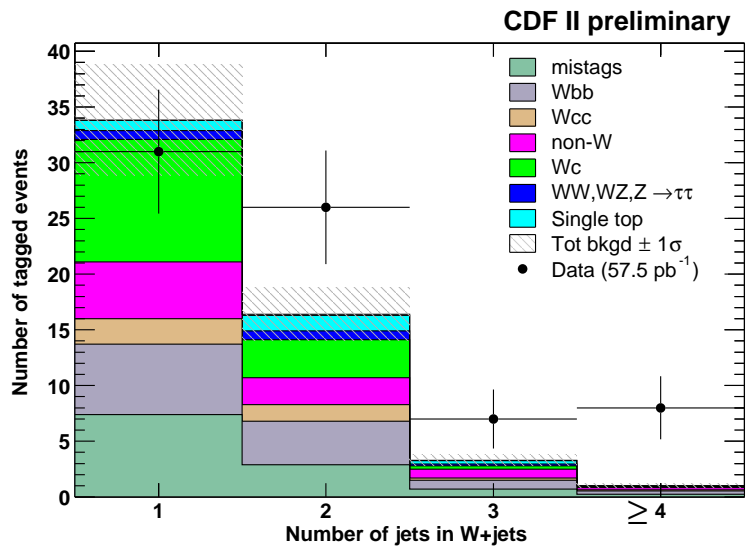}
}
\end{center}
\caption{$W$+N jets spectrum of the secondary vertex $t\bar{t}$ analysis 
         at CDF. The 15 candidate events are shown as dots. The various
         backgrounds are shown as shaded areas in different colors.
         The access of events in the $W$+3 and 4 jets bin is interpreted
         as $t\bar{t}$ signal.}
\label{fig:CDF_WNjets}       % Give a unique label
\end{figure}

In this analysis one uses the fact that only about 2\% of generic
$W$+jets events contain real b-quarks. If b-quarks can be identified
with good efficiency and purity, the $W$ + multi-jet background
can be considerably suppressed.
The preselection of this analysis calls for one tightly identified 
electron or muon in the central detector. An isolation requirement
is applied. Conversion, Z-boson and cosmic ray vetos are imposed.
The missing transverse energy has to be larger than 20~GeV.
The b-tagger program selects good tracks and tries to
form secondary vertices. The 2D decay length in the $x$-$y$
plane is calculated and taking the vertex resolution into account
a vertex significane is formed. 
The version of the tagger used for the Winter 2003 analysis
presented here used the beamline to estimate the position
of the primary interaction vertex and yielded an
event tagging efficiency for $t\bar{t}$ events
of $45\%\pm1\%(\mathrm{stat})\pm 5\%(\mathrm{syst})$.
At least one b-tagged jet is required in this analysis.
The signal region is defined as the $W$+3 and $W$+4 jets 
bin. The jet definition is $E_T>15$~GeV and $|\eta|<2.0$.
CDF finds 15 candidate events over a background
of 4.1 events in 57.5~$\mathrm{pb^{-1}}$ of data.
The cross-section is found to be:
\[\sigma(t\bar{t}) = 5.3\pm1.9(\mathrm{stat})\pm0.8(\mathrm{syst})
  \pm0.3(\mathrm{lumi})\;\mathrm{pb}. \]
The $W$+N jets spectrum is shown in Fig.~\ref{fig:CDF_WNjets}.

\subsection{The Soft Muon Tag at $\DZeroBf$}
The soft muon tagger searches for muons coming from
semileptonic b-decays in jets. The muons must have a minimum
$p_T > 4$~GeV and $|\eta|<2.0$. The distance in the 
$\eta$-$\phi$ plane must be smaller than 
$\Delta R (\mu, \mathrm{jet})<0.5$.
$\DZero$ observes 2 events in this analysis over an expected
background of $0.9\pm0.39$ events.
The cross-section combined with the topological analysis
is 
\[ \sigma(t\bar{t}) = 5.8^{+4.3}_{-3.4}(\mathrm{stat})\,^{+4.1}_{-2.6}
   (\mathrm{syst})\pm0.6(\mathrm{lumi})\;\mathrm{pb}\]

\section{Conclusions}
Top quark signals have been reestablished at the Tevatron in the di-lepton
and the lepton+jets channels. Various analysis techniques are used
and yield results which are in agreement with each other and the 
theoretical prediction.
Data sets larger than those of Run I are at hand and will in the 
near future allow cross-section measurements well below
Run I uncertainties. The final goal for Run IIa 
(2~$\mathrm{fb^{-1}}$ of data) is to achieve a relative uncertainty of 7\%. 

%
% BibTeX users please use
% \bibliographystyle{}
% \bibliography{}

\begin{thebibliography}{}
%
% and use \bibitem to create references.
%
\bibitem{maruyama}Takasumi Maruyama, Top Quark Mass at CDF.
  To appear in the proceedings of International Europhysics Conference on 
  High-Energy Physics (HEP 2003), Aachen, Germany, 17-23 Jul 2003.
% Format for Journal Reference
% Author, Journal \textbf{Volume}, (year) page numbers.
% Format for books
\bibitem{merkel} Petra Merkel, 
  Search for Physics beyond the standard model in $t\bar{t}$ Production.
  To appear in the proceedings of International Europhysics Conference on 
  High-Energy Physics (HEP 2003), Aachen, Germany, 17-23 Jul 2003.

%Author, \textit{Book title} (Publisher, place year) page numbers
% etc
\bibitem{shabalina} Elizaveta Shabalina, Top Quark production at the Tevatron.
  FERMILAB-CONF-03-317-E. 
  To appear in the proceedings of International Europhysics Conference on 
  High-Energy Physics (HEP 2003), Aachen, Germany, 17-23 Jul 2003.

\bibitem{warsinsky} Markus Warsinsky, Top Quark Physics with D0.
  To appear in the proceedings of International Europhysics Conference on 
  High-Energy Physics (HEP 2003), Aachen, Germany, 17-23 Jul 2003.
\end{thebibliography}
%
% Non-BibTeX users please use

\end{document}